# Manipulate Elastic Wave Modes by an Ultrathin Three-component Elastic Metasurface

Pai Peng*, Cheng Feng and Kangcheng Zhou

School of Mathematics and Physics, China University of Geosciences, Wuhan 430074, China

**Abstract**

We design a two-dimensional ultra-thin elastic metasurface consisting of steel cores coated with elliptical rubbers embedded in epoxy matrix, capable of manipulating bulk elastic wave modes for reflected waves. The energy exchanges between the longitudinal and transverse modes are completely controlled by the inclined angle of rubber. One elastic mode can totally convert into another by the ultra-thin elastic metasurface. The conversion mechanism based on the non-degenerate dipolar resonance is a general method and easily extended to three-dimensional or mechanical systems. A mass-spring model is proposed and well describe the conversion properties. We further demonstrate that high conversion rates (more than 95%) can be achieved steadily for one elastic metasurface working on almost all different solid backgrounds. It will bring wide potential applications in elastic devices.

E-mail: paipeng@cug.edu.cn

In the past decade, the emerging of metasurfaces [1], which are ultrathin material slabs with sub-wavelength artificial structures providing gradient phase delay on the surface, make it possible to modulate electromagnetic wavefront. The generalized Snell's law opens a new degree of freedom in manipulating electromagnetic waves. This concept of metasurfaces has been extended to acoustic metasurfaces. Full control (phase and amplitude) of scalar acoustic waves by acoustic metasurfaces has attracted a lot of research interests [2-10]. Various of acoustic metasurface based on coiling up structures [2-4] or Helmholtz resonators [5] are proposed to modulate acoustic wavefront. Very recently, Ghaffarivardavagh et al. [7] and Zhu et al. [8] present acoustic metasurfaces enabling acoustic control with simultaneous phase and amplitude modulation [9]. These works on amplitude manipulation fill the gap in full control of acoustic waves.

Elastic waves exhibit rich polarization catachrestic absent in acoustic and electromagnetic waves. Beside the phase and amplitude, polarization modulation is another challenging project specially for elastic waves. The concept of metasurface provide a new way for modes conversion of elastic waves. Until now, the modulating of elastic wave wavefront by elastic metasurfaces is widely studied in solid slabs [11-16]. Very few works are carried out on bulk elastic waves [17-20]. To the best of our knowledge, it is still a blank in of amplitude or polarization properties manipulation for elastic waves. Some efforts have been made on interesting mode-coupled wave behavior in solids [20-23]. However, critical incident angles or anisotropic materials are used in achieving total wave conversions. The manipulation of elastic wave modes by elastic metasurfaces has not yet been explored.

In this letter, we design an ultra-thin elastic metasurface based on "three-component" resonators [24,25], which feature the elastic metasurface at low frequency. The elastic metasurface brings strong couplings between L and T modes, resulting in high mode conversions. We focus on the conversions from L-to-T modes, because it is harder than the conversion from T-to-L modes, and T waves have great potential in medical and industrial applications due to its short wavelength [21].

We consider an elastic metasurface placed on the surface of a semi-infinite epoxy background, as shown in Figs. 1(a). The elastic metasurface consists of periodic steel cylinders coated with elliptical soft rubbers embedded in epoxy matrix, as shown in Figs. 1(b). The radius of the steel is $r = 0.2p$, where $p$ is the periodic constant. The semimajor and semiminor axes of the elliptical rubber layer are $r_a = 0.38p$ and $r_b = 0.24p$, respectively, and $\theta$ is the included angle between the major axes of ellipse and the horizontal direction (x-axes). The parameters of materials used are: $\rho_e = 1180 \text{kg/m}^3$, $c_{l,e} = 2540 \text{m/s}$ and $c_{t,e} = 1160 \text{m/s}$ for epoxy; $\rho_r = 1300 \text{kg/m}^3$, $c_{l,r} = 55 \text{m/s}$ and $c_{t,r} = 22 \text{m/s}$ for rubber [24]; $\rho_s = 7900 \text{kg/m}^3$, $c_{l,s} = 5800 \text{m/s}$ and $c_{t,s} = 3200 \text{m/s}$ for steel, respectively, where $\rho$ is the mass density, and $c_l$ and $c_t$ are L and T wave speeds.

We let plane L waves normally incident from bottom, and both L and T modes are found in reflected waves. That means after reflecting by the elastic metasurface, some incident L waves have been converted into T waves. The L-to-T modes conversion rates (CR) can be described by the ratio $E_R^T / E_I^L$ of their energy flux along the y-direction:

$$\frac{E_R^T}{E_I^L} = \frac{\left| \int_0^p \left( -v_x^* \right) \cdot \tau_{xy} dx \right|}{\left| \int_0^p \left( -v_y^* \right) \cdot \tau_{yy} dx \right|} \quad (1),$$

where $v$ and $\tau$ are the velocity (symbol * means conjugate) and stress on the interface between the elastic metasurface and the epoxy background, and scripts R (I) and T (L) denote the reflected

(input) wave and transverse (longitudinal) mode, respectively. The CR as a function of frequency $\omega$ and the included angle $\theta$ is computed by COMSOL Multiphysics software based on the finite-element method. As shown in Figs. 1(c), the value of CR is generally very small due to the weak coupling between L and T modes. Large CEs are found near two frequencies $\omega_a \approx 0.0083(2\pi c_{t,e}/p)$ or $\omega_b \approx 0.1151(2\pi c_{t,e}/p)$. In particular, CR can reach its theoretical maximum value ($E_R^T/E_I^L \leq 1$ due to the conservation of energy) at a best-angle $\theta_a = 0.31(\pi)$ or $\theta_b = 0.19(\pi)$ for the frequency $\omega_a$ or $\omega_b$, respectively. That means there are no L waves in reflection. The incident L mode elastic waves have been totally converted into T mode in reflected waves. The two white thick arrows shown in Figs. 1(c) point out where the total conversion effect occurs. We note that the wavelength in epoxy background is about 2 orders higher than the thickness of the elastic metasurface.

We further calculate the band structure of an elastic metamaterial made of the same square unit cells as the elastic metasurface. The included angle is $\theta = \theta_a$. The band structure of the elastic metamaterial is plotted in Figs. 2(a). The lowest flat band at frequency $\omega \approx 0.0065(2\pi c_{t,e}/p)$ is induced by locally rotational resonance [26]. The second and third branch are actually induced by a non-degenerate dipolar resonance [25]. The displacement field distributions of the eigenstates of the eigenmodes on the second and third branch at the X point of the Brillouin zone are plotted in Figs. 2(c) and (b), respectively. The movements of the cores in these two eigenstates are perpendicular to each other. The eigenfrequencies of the two non-degenerate eigenmodes are $\omega_a \approx 0.0083(2\pi c_{t,e}/p)$ and $\omega_b \approx 0.1151(2\pi c_{t,e}/p)$, which are exactly the frequencies at where the total conversion is obtained. These two eigenfrequencies are almost independent with the included angle $\theta$.

The conversion mechanism can be explained with a simple picture. In intuitively, the high conversion must rely on two factors: the elliptical rubber layer and the resonance inside the elastic metasurface. The inclined elliptical rubber layer breaks the system symmetry and makes the L and T modes coupled to each other. The coupling is weak unless the resonance is involved. When frequency is near to the eigenfrequencies $\omega_a$ or $\omega_b$, the excited dipolar resonance brings very strong oblique vibrations inside the elastic metasurface. To satisfy the continuous boundary condition on the interface between the elastic metasurface and the epoxy background, T modes is required to exist in reflected waves if horizontal movements are absent in incident waves. As a result, L-to-T modes conversion is observed. The T-to-L mode conversion can be obtained in a similar way. The design of dipolar resonance induced modes conversion is easily extended to three-dimensional or mechanical systems.

We propose a mass-spring model, as shown in Figs. 3(a), to captures the main physical essence of the conversion effect and reveals the relation between the CR and the included angle $\theta$. The model consists of a resonator [25] connects to a half one-dimensional monatomic chain. The elastic metasurface is modeled as the resonator. The core and matrix in one unit cell are modeled as mass $m$ (denoted by red disk) and $M$ (denoted by green ring), respectively. The mass $m$ is connected by four springs to the mass $M$. The stiffnesses of the two springs, which is induced by the deformations inside the rubber layer, along the major axes direction of the ellipse are given by $L_a$ and $G_a$ for extensional and shear displacements [25], respectively. Similarly, the stiffness of the two springs along the minor axes direction of the ellipse are given by $L_b$ and $G_b$, respectively. The horizontal (vertical) displacements of mass $m$ and $M$ are $u$ ($v$) and $U$ ($V$), respectively. The expressions of the forces $F_{m,a}$ and $F_{m,b}$ exerted on $m$ along the major and

minor axes directions of the ellipse, respectively, are given as following:

$$\begin{cases} F_{m,a} = -2L_a\left[(u-U)\cos\theta + (v-V)\sin\theta\right] - 2G_b\left[(u-U)\cos\theta + (v-V)\sin\theta\right] \\ F_{m,b} = -2L_b\left[(v-V)\cos\theta - (u-U)\sin\theta\right] - 2G_a\left[(v-V)\cos\theta - (u-U)\sin\theta\right] \end{cases} \quad (2).$$

As plane L waves are normally incident to the elastic metasurface, the semi-infinite solid background is modeled as a half one-dimensional monatomic chain. The mass of the chain is $m_0$ related to the mass density of the epoxy background, and the spring stiffness of the chain is $L$ and $G$ related to the compressional and shear modulus of the epoxy background, respectively. The distance between neighbor masses $m_0$ is denoted as $d$. The horizontal and vertical displacement of the $n$-th mass $m_0$ is denoted as $u_n$ and $v_n$, respectively. We apply the assumptions of harmonic time dependence $e^{-i\omega t}$ and the Bloch's condition $v_{n+1} = v_n e^{iq_L d}$ ($u_{n+1} = u_n e^{iq_T d}$) to the chain, where $q_L = \omega/c_{l,e}$ ($q_T = \omega/c_{t,e}$) is the Bloch wave vector for longitudinal (transverse) waves. The displacements on the 0-th and 1-th mass $m_0$ can be written as:

$$\begin{cases} v_0 = A_I + A_R \\ u_0 = B_I + B_R \\ v_{-1} = A_I e^{-iq_L d} + A_R e^{iq_L d} \\ u_{-1} = B_I e^{-iq_T d} + B_R e^{iq_T d} \end{cases} \quad (3).$$

Here $A_I$ ($B_I$) and $A_R$ ($B_R$) are arbitrary amplitude for upward and downward L (T) waves. The 0-th mass $m_0$ is connected to the mass $M$ by a spring the same to the springs between the masses $m_0$. According to Newton's second law, we can get the equations of motion in global coordinate:

$$\begin{cases} m\ddot{u} = F_{m,a}\cos\theta - F_{m,b}\sin\theta \\ m\ddot{v} = F_{m,a}\sin\theta + F_{m,b}\cos\theta \\ M\ddot{U} = -G(U-u_0) - F_{m,a}\cos\theta + F_{m,b}\sin\theta \\ M\ddot{V} = -L(V-v_0) - F_{m,a}\sin\theta - F_{m,b}\cos\theta \\ m_0\ddot{u}_0 = -G(u_0-U) - G(u_0-u_{-1}) \\ m_0\ddot{v}_0 = -L(v_0-V) - L(v_0-v_{-1}) \end{cases} \quad (4).$$

In case of L wave incidence, we have $B_I = 0$. By using Eqs. (2) and Eqs. (3) into Eqs. (4), there are six equations for six unknowns as $u$, $v$, $U$, $V$, $A_R$, and $B_R$. The expression of CR can be obtained from the model as $E_R^T / E_I^L = (c_{t,e} / c_{l,e})|B_R / A_I|^2$. The analytical results of the CR are plotted in Figs. 3(b), which is agreed well with the simulation results shown in Figs. 2(c). The total conversion effects are obtained at the same frequencies and the same included angles. The parameters used in the model are $m = \rho_s \pi r^2$, $M = \rho_e(p^2 - \pi r_a r_b)$, $m_0 = \rho_e d^2$, $\omega_a^2 = 2(L_a + G_b)/m$, $\omega_b^2 = 2(L_b + G_a)/m$, $L = \rho_e c_{l,e}^2$, $G = \rho_e c_{t,e}^2$ and $d = p$, where the value of $\omega_a$ and $\omega_b$ are obtained from the band structure shown in Figs. 2(a).

The expression of CR can be greatly reduced thus giving a clear picture for the conversion effect under two approximations: "low frequency" and "soft rubber". The studied system (shown in Figs. 1(a)) is physically qualified to meet these two approximations. As the dipolar resonance are set at very low frequency, resulting from heavy cores and soft coating layers, $\omega$ can be regarded as an infinitely small quantity. As the rubber is much softer than the other materials, the terms with modulus ratio between the rubber and other materials $L_a/L$ (or $G_a/G$, etc.) are also regarded as infinitely small quantities. After ignoring the high-order terms of the small quantities, the expression of CR can be rewritten as:

$$\frac{E_R^T}{E_I^L} = \frac{c_{t,e}}{c_{l,e}} \left| \frac{\sin 2\theta \left(\omega_b^2 - \omega_a^2\right)}{\left(\omega_a^2 \cos^2\theta + \omega_b^2 \sin^2\theta - \frac{\omega_a^2 \omega_b^2}{\omega^2}\right) + \frac{c_{t,e}}{c_{l,e}} \left(\omega_a^2 \sin^2\theta + \omega_b^2 \cos^2\theta - \frac{\omega_a^2 \omega_b^2}{\omega^2}\right)} \right|^2 \quad (5).$$

Equation (5) shows that the value of CR strictly equal to zero in two conditions: $\sin 2\theta = 0$ or $\omega_a - \omega_b = 0$. The first one means the elliptical rubber layer is orthogonally placed, and the second one means rubber layer is circle not ellipse. Both these two conditions lead to decoupling between the L and T modes. To obtain high CE, we let the frequency $\omega$ equal to $\omega_a$, then the Eq. (5) is reduced to

$$\frac{E_R^T}{E_I^L} = \frac{c_{t,e}}{c_{l,e}} \left| \frac{-2c_{l,e}}{c_{t,e} \tan\theta + c_{l,e} \cot\theta} \right|^2 \quad (6).$$

At the resonance frequency $\omega_a$, the CR is as a function of the included angle $\theta$. Inequality operating on the denominator on the right side of the equation, Eq. (6) is rewritten to

$$\frac{E_R^T}{E_I^L} \leq \frac{c_{t,e}}{c_{l,e}} \left| \frac{-2c_{l,e}}{2\sqrt{c_{t,e}c_{l,e}}} \right|^2 = 1 \quad (7).$$

The inequality sign shows that the value of CR can reach to one *iff* $c_{t,e} \tan\theta = c_{l,e} \cot\theta$, which gives the expression of best-angle $\theta_a = \arctan\sqrt{c_{l,e}/c_{t,e}}$. To our surprise, the best-angle $\theta_a$ is independent with the elastic metasurface but simply determined by the ratio between L and T wave speeds (or say the Poisson's ratio) in the solid background. We plot the analytical results (red line) of CR obtained from Eq. (6), which perfectly predict the simulated results (black dots), showing in Figs. 4(a). The reflected wave will be pure L wave when $\theta$ equals to 0 or $\pi/2$, where no conversion happens. On the contrary, the reflected wave will be pure T wave when the $\theta$ equals to $\theta_a$, where the total conversion happens. Otherwise, both L and T modes will be found in reflected waves. The compositions of elastic modes in reflected waves are controllable independently by change the included angle $\theta$. Similarly, by using $\omega = \omega_b$ into Eq. (5), we get

$\theta_b = \arctan \sqrt{c_{t,e} / c_{l,e}} = \pi - \theta_a$. As shown in Figs. 4(a), the simulated results (blue squares) are also perfectly predicted by the model (green line).

Another interesting phenomenon is that the high conversion effect is very steady for different background materials. As shown in Figs. 4(a) red line, high CR (defined as $E_R^T / E_I^L \geq 0.95$) is achieved in a range of $\theta$ from $0.28\pi$ to $0.35\pi$. The corresponding range of the ratio between L and T wave speeds is from $\sqrt{2}$ to 3.69 (Poisson's ratio from 0 to 0.46). Actually, it is an extremely wide range covered almost all the solid materials. For example, when the epoxy background is replaced by beryllium, which is a common material with large difference between L and T wave speeds, the change of the best-angle is very small. For a fixed included angle $\theta = \theta_a$, the CR will only slightly reduce from 1 to 0.98 after changing the background, as shown in Figs. 4(b). For other common solids such as aluminum, glass, or lead instead of beryllium, the CEs keep higher than 0.99. We can build an elastic metasurface and use to stick to different solid if high conversion is always requested in practical applications. The steadiness property greatly enhances the applicability of the elastic metasurface in potential applications.

In conclusion, we design an ultra-thin elastic metasurface with ability in manipulating reflected elastic wave modes. The elastic metasurface contains obliquely placed elliptical coating layers, which break the decoupling between L and T elastic modes and bring strong non-degenerate dipolar resonance. Input elastic waves can be totally converted into the other mode after reflected by the elastic metasurface. The total conversion effect occurs with special frequency and included angle. The special frequency is the eigenfrequency of the dipolar resonance, and the special included angle depends on but insensitive to the ratio between the L and T wave speeds in background. High conversion could always be obtained for different background materials, which brings potential applications in elastic devices.

The authors would like to thank Professor Z. Y. Liu for discussions. This work was supported by the National Natural Science Foundation of China (Grants No. 1160040476).

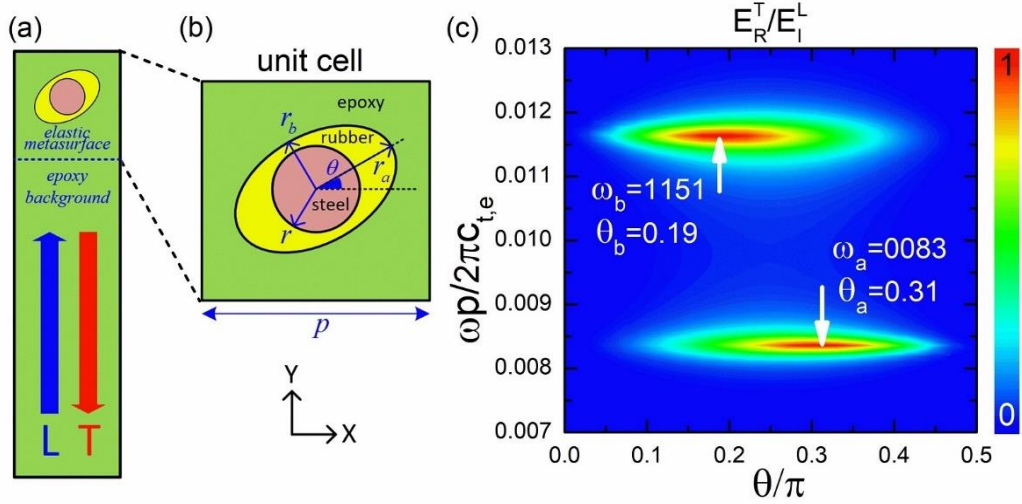

FIG. 1. (a) Schematic of the unit cell of our periodic system along the *x*-direction. Periodic boundary conditions are applied on the left and right side, free boundary condition is applied on the top side, and absorbing boundary condition is applied on the bottom side. Plane longitudinal waves are normally incident from the bottom, and reflected waves are transverse waves. (b) Schematic of a zoom view for one unit cell of the elastic metasurface. Here $r$, $r_a$ and $r_b$ are the radii for steel core, semimajor and semiminor axes of the rubber layer, respectively. $p$ and $\theta$ are the period constant and included angle. (c) Simulation results of the converting efficiency $E_R^T / E_I^L$ as a function of frequency $\omega$ and the included angle $\theta$. Thick white arrows indicate where the total conversion effect ($E_R^T / E_I^L = 1$) occurs.

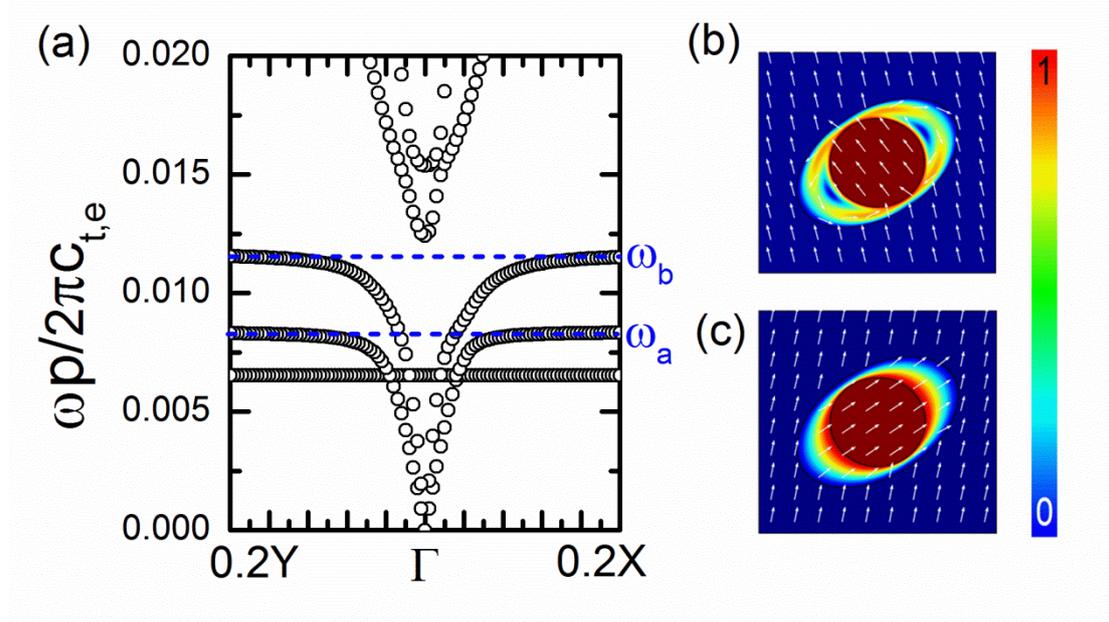

FIG. 2. (a) The band structure of elastic metamaterials consisting of the same unit cells as the elastic metasurface. The included angle is $\theta = \theta_a$. Blue dashed lines denote the two eigenfrequencies $\omega_a$ and $\omega_b$ for a non-degenerate dipolar resonance. (c) and (b) are the displacement field distributions of eigenmodes on the second and third band at the X points, respectively. Dark red and dark blue correspond to 1 and 0 of the normalized magnitude, respectively, and thin white arrows indicate directions.

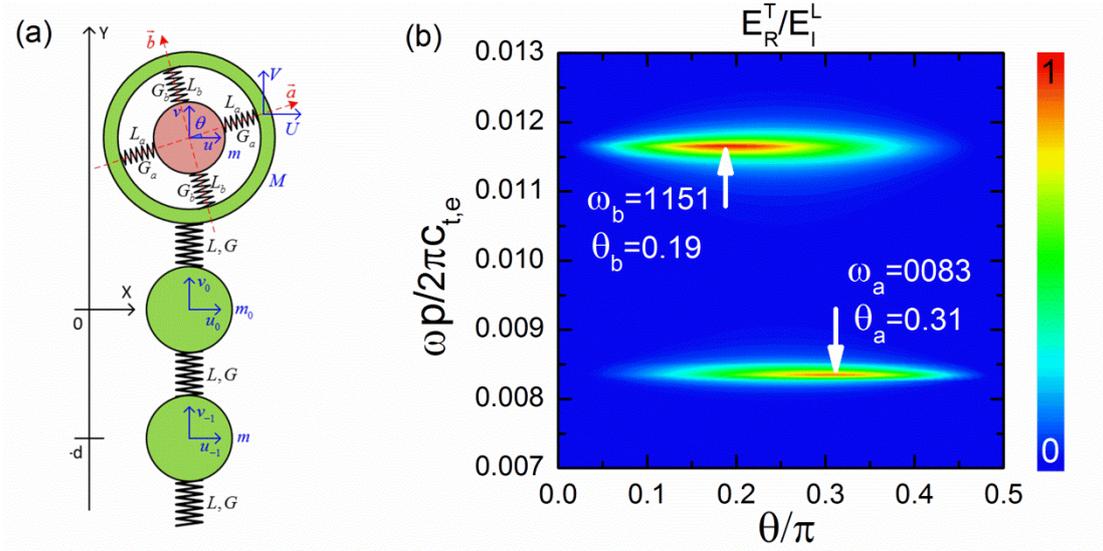

FIG. 3. (a) A schematic view of the mass-spring model. The light red disk and the greed ring indicates the steel with mass $m$ and the matrix with mass $M$, respectively, in one unit cell of the elastic metasurface. The spring stiffness along the major axes for extensional and shear displacements are $L_a$ and $G_a$, those along the major axes are $L_b$ and $G_b$, respectively. The half one-dimensional monatomic chain below $M$ indicate the solid background with mass $m$ and spring stiffness $L$ and $G$. (b) Analytical results obtained from the mass-spring model. The converting efficiency $E_R^T / E_I^L$ as a function of frequency $\omega$ and the included angle $\theta$. Thick white arrows indicate where $E_R^T / E_I^L = 1$.

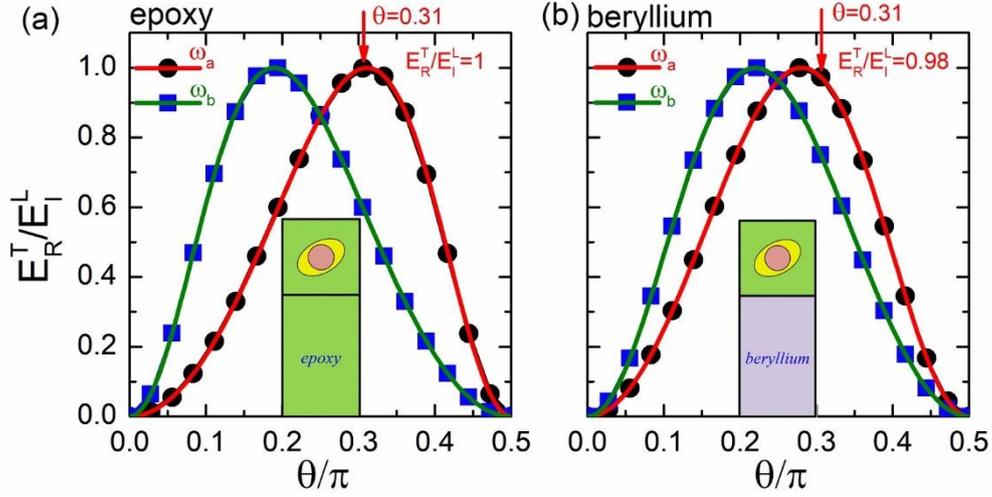

FIG. 4. (a) The converting efficiency is as a function of the angle $\theta$ with a given frequency $\omega_a$ or $\omega_b$. The red line and black dots indicate the analytical results obtained from the Eq. (6) and the simulation results shown in Figs. 1(c), respectively. The green line and blue squares indicate those results with a given frequency $\omega_b$. (b) is the same with (a) but a different background of beryllium. The used material parameters for beryllium are $\rho_b = 1850 \text{kg/m}^3$, $c_{l,b} = 12800 \text{m/s}$ and $c_{t,b} = 7800 \text{m/s}$.